\documentclass[showpacs,aps,twocolumn]{revtex4}

\usepackage{bm}
\usepackage{amsmath}
\usepackage{graphicx}
\usepackage{subfigure}
\usepackage[usenames,dvipsnames]{color}
\definecolor{darkblue}{RGB}{0,0,196}
\usepackage[colorlinks=true,linkcolor=darkblue,citecolor=darkblue,urlcolor=darkblue]{hyperref}

\usepackage{xcolor}

\usepackage{setspace}
\usepackage{footmisc}
\usepackage[makeroom]{cancel}
\usepackage{comment}
\usepackage{lineno}

\def\be{\begin{equation}}
\def\ee{\end{equation}}
\def\ba{\begin{eqnarray}}
\def\ea{\end{eqnarray}}

\usepackage{graphicx}
\usepackage{amsmath,bbm}
\usepackage{amssymb,bm}
\usepackage{yfonts}

\begin{document}

\title{Identified particle production in Xe+Xe collisions at $\sqrt{s_{\rm{NN}}}$ = 5.44 TeV using a multiphase transport model}
\author{Rutuparna Rath}
\author{Sushanta Tripathy}
\author{Raghunath Sahoo\footnote{Corresponding author: $Raghunath.Sahoo@cern.ch$}}
\author{Sudipan De\footnote{Presently at NISER, Bhubaneswar}}
\affiliation{Discipline of Physics, School of Basic Sciences, Indian Institute of Technology Indore, Simrol, Indore 453552, India}

\author{Mohammed Younus}
\affiliation{Department of Physics, Nelson Mandela University, Port Elizabeth, 6031, South Africa}

\begin{abstract}
\noindent
Xe+Xe collisions at relativistic energies provide us with an opportunity to study a possible system with deconfined quarks and gluons, whose size is in between those produced by p+p and Pb+Pb collisions. In the present work, we have used AMPT transport model with nuclear deformation to study the identified particle production such as ($\pi^{+}+\pi^{-}$), (K$^{+}$+K$^{-}$), $\rm{K}_{s}^0$, (p+$\bar{\rm{p}}$), $\phi$ and ($\Lambda + \bar{\Lambda}$) in Xe+Xe collisions at $\sqrt{s_{\rm NN}}$=5.44 TeV. We study the $p\rm{_T}$-spectra, integrated yield, $p\rm{_T}$-differential and $p\rm{_T}$-integrated particle ratios to ($\pi^{+}+\pi^{-}$) and (K$^{+}$+K$^{-}$) as a function of collision centrality. The particle ratios are focused on strange to non-strange ratios and baryon to meson ratios. The effect of deformations has also been highlighted by comparing our results to non-deformation case. We have also compared the results from AMPT string melting and AMPT default version to explore possible effects of coalescence mechanism.  We observe that the differential particle ratios show strong dependence with centrality while the integrated particle ratios show no centrality dependence.
We give thermal model estimation of chemical freeze-out temperature and the Boltzmann-Gibbs Blast Wave analysis of kinetic freeze-out temperature and collective radial flow in Xe+Xe collisions at $\sqrt{s_{\rm{NN}}}$ = 5.44 TeV. 

\end{abstract}
 
\pacs{12.38.Mh, 25.75.Ld, 25.75.Dw}
\date{\today}
\maketitle 
\section{Introduction}
\label{intro}
Ultra-relativistic heavy ion collisions experiments conducted at RHIC and LHC give us opportunities to peek into the past when Universe was a few microseconds old. The collisions result into a system of deconfined quarks and gluons at very high temperature and density or, quark-gluon-plasma (QGP)~\cite{QGP}. Till recent times, mainly symmetrical nuclei such as lead (Pb) ions or assuming spherical gold (Au) ions have been used to collide and form QGP. Recently, interests have come forth to conduct experiments with intrinsically deformed nuclei. Experiments have been conducted at RHIC, BNL with Uranium (U), which is heavier than gold and lead ions and is considered to be highly deformed (lead ion has zero deformity). A comparison of central collision of spherical nuclei with that of deformed nuclei helps in establishing if the elliptic flow observed in heavy-ion collisions, which is considered as a signature of QGP, is an initial state effect \cite{v2,v2_star1,Adamczyk:2015obl}. In case of a deformed nuclei collision, one expects the charged particle multiplicity density in the transverse phase space to be higher as compared to the collision of spherical nuclei \cite{Nepali:2006ep,mws_2,Tripathy:2018rln}. Particle density per unit volume in ideal hydrodynamical models is independent of mass number of colliding species. A violation of scaling behavior is seen in the observables due to the deformed structures of the colliding nuclei \cite{Giacalone:2017dud}.
\begin{figure}[ht]
\includegraphics[height=17em]{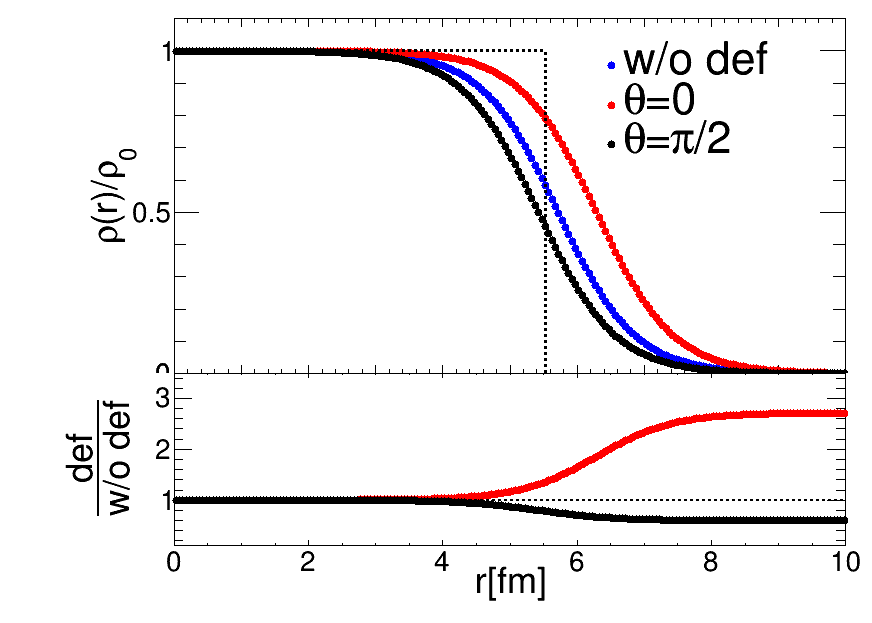}
\caption[]{(Color online) The nuclear density profile for Xenon nucleus. Shown are the hard sphere, Woods-Saxon and deformed Woods-Saxon density profiles. The bottom panel shows the ratio of nuclear deformations with respect to no-deformation for Xenon nucleus.}
\label{fig0}
\end{figure}

The LHC has collided Xenon ($^{129}\rm Xe$) ions at $\sqrt{s_{\rm NN}}$ = 5.44 TeV to bridge the final state multiplicity gap between the larger Pb-ion systems and smaller systems like p+p and p+Pb. The new findings at the LHC shows that the identified particle production as a function of normalized charged particle multiplicity is independent of collision species and collision energy \cite{Tripathy:2018ehz}. Eventually, the final state multiplicity density of the system drives the dynamics of particle production. In view of this, Xe+Xe collisions serves to bridge the multiplicity gap between p+p, p+Pb to Pb+Pb collisions and to help in the observation of a universal scaling. It is also observed at the LHC that Xe+Xe collision system violates the quark participant scaling of charged particle production, similar to other collision species with spherical nuclei \cite{v2_star2,v2_phenix3,v2_phenix4,AMPT_Scaling,AMPT_PbPb,Kim:2018ink}. In addition, Xe being a deformed nucleus would help in understanding many new features like those observed at the RHIC using U+U collisions, but at a much higher collision energy. It has been shown recently that intrinsic deformities may effect particle flows for central collisions while for peripheral collisions this effect is negligible ~\cite{Acharya:2018ihu}.

Let us now briefly discuss the particle production in relativistic heavy ion collisions.  Heavy ion collisions produce a system of deconfined quarks and gluons for infinitesimally small time, and soon disintegrate by forming hadrons. While initially we have only nucleons and their up and down quarks within the nuclei interacting during collisions, almost all types of known hadrons (including nucleons) are finally detected and this indicates that all six types of quarks and much more gluons in extra are produced (also photons, leptons) which were absent initially. Study of such enhancement in particle density in comparison to initial ground state nuclear density gives us the direct proof of such high temperature and dense state of quarks and gluons called QGP. Particle ratios mainly ratios of different hadrons give us a direct picture of such enhancement of different quarks and also how they behave and interact as part of the bulk medium~\cite{Abelev:2014uua}. In particular, it is believed that medium flow~\cite{AMPT1,AMPT3,v2_star4} from the point of the collision as well as any form of fluctuation or anisotropy in the initial stages of collision may greatly determine the spectral shape and nature of these ratios. In this context, one may be tempted to note that the momentum anisotropies which are the results of nuclei colliding at different impact parameters may affect the particle ratios. This particular aspect is currently being investigated in many experimental and theoretical studies. As mentioned earlier in the introduction, till recent times, we have used either Au or Pb nuclei to investigate the issues, which have either assumed spherical shape or zero deformities. However, it becomes desirable to explore the effect of nuclear 
deformation on particle production and its effect on properties of matter, which are sensitive to nuclear geometry.
In this paper we have introduced non-zero deformities to the Xe nuclei to study effects on the particle production. Collisions of Xenon nuclei should provide us with a much cleaner system than Pb nuclei as well as a more denser and a hot QGP medium than proton collisions, p+p might form at comparable collision energies. In addition to this, we have effects due to deformations which we may be able to discern with much less efforts than in case of U+U collisions. 

In this paper we have included deformation to the nucleus defined within AMPT model. We will discuss this briefly in one of the following sections. We have calculated particle ratios for the charged hadrons and have tried to find out the effects of the deformation on particle production. The paper is organized as follows. The present section of introduction is followed by sections on formalism and results and discussion respectively. These are followed by conclusion at the end.
\begin{figure}[ht]
\includegraphics[height=17em]{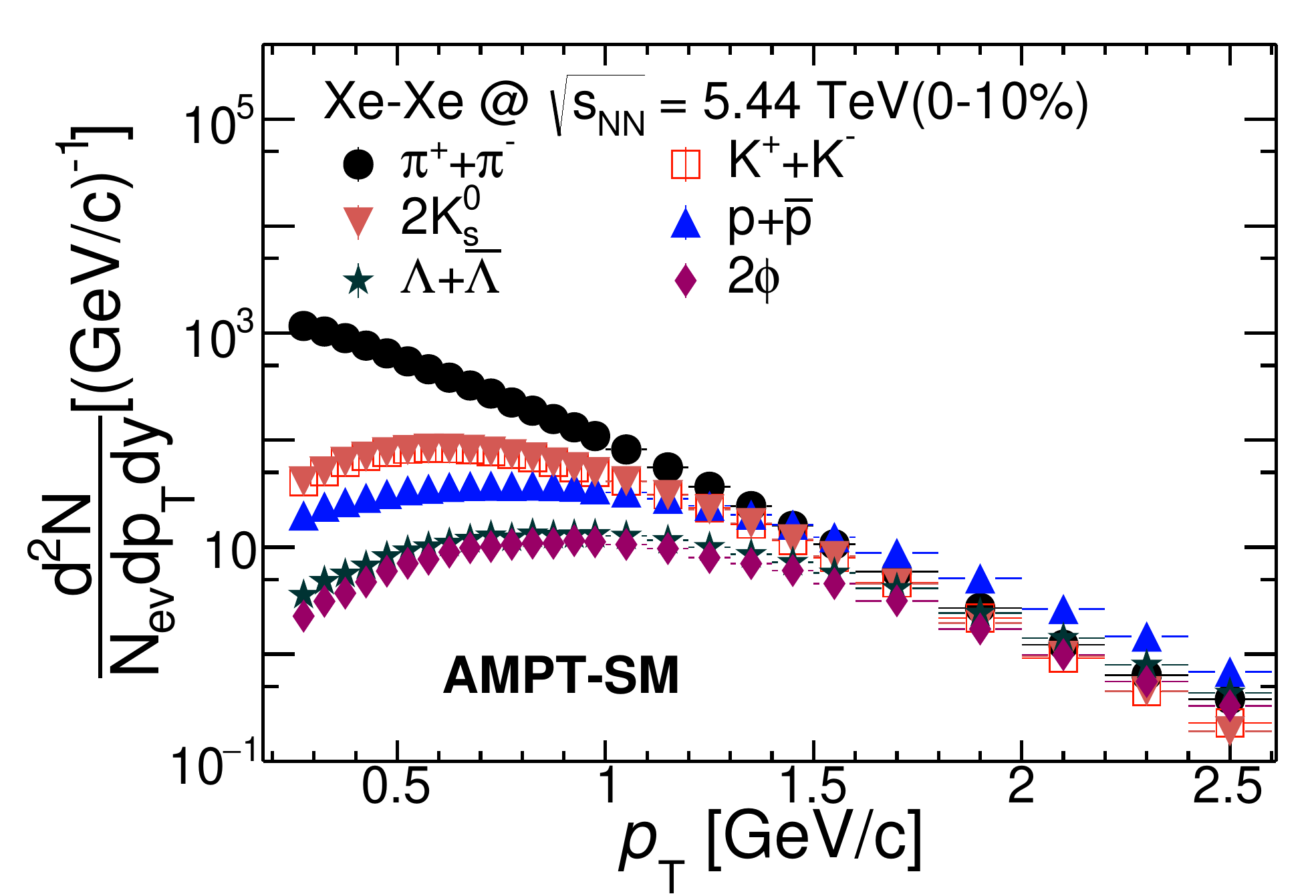}
\caption[]{(Color online) $p\rm{_T}$-spectra of identified particles in Xe+Xe collisions at $\sqrt{s_{\rm NN}}$ = 5.44 TeV for 0-10\% centrality using AMPT-SM. Different symbols show different particle species. The vertical lines in the results show the statistical uncertainties.}
\label{fig1}
\end{figure}

\section{Formalism}
\label{formalism}
\subsection*{A Mutli-Phase Transport (AMPT) model}
\label{ampt}
AMPT is a hybrid transport model which contains four components namely, initialization of collisions, parton transport after initialization, hadronization mechanism and hadron transport~\cite{AMPT2}.  The initialization of the model follows HIJING model~\cite{ampthijing} and calculates the differential cross-section of the produced minijet particles in p+p collisions which is given by,

\begin{eqnarray}
\frac{d\sigma}{dp_T^2\,dy_1\,dy_2}=&K&\,\sum_{a,b}x_1f_a(x_1,p_{T1}^2)\,x_2f_2(x_2,p_{T2}^2)\nonumber\\
&\times&\frac{d\hat{\sigma}_{ab}}{d\hat{t}}\,,
\end{eqnarray}
where $\sigma$ is the produced particles cross-section and $\hat{t}$ is the momentum transfer during partonic interactions in p+p collisions. $x_i$'s are the momentum fraction of the mother protons which are carried by interacting partons and $f(x, p_T^2)$'s are the parton density functions (PDF). 
The produced partons calculated in $\rm p+p$ collisions is then converted into $\rm A+A$ and $\rm p+A$ collisions by incorporating parametrized shadowing function and nuclear overlap function using in-built Glauber model within HIJING.
In case of Pb nucleus, we use Woods-Saxon (WS)~\cite{Loizides:2017ack} distribution to define the distribution of nucleons (HIJING). For the deformed nucleus such as Xenon, we may include deformation parameter, $\beta_n$, along with spherical harmonics, $Y_{nl}(\theta)$, in the WS function~\cite{mws_1,mws_2, mws_3,mws_4,mws_5}. This is known as modified Woods-Saxon (MWS) density distribution. We have used MWS within the HIJING model to calculate initial distributions of partons etc., for tip, body or random configuration collisions of Xenon nuclei. Let us now describe briefly MWS.
Nucleon density in HIJING is usually written as a three parameter Fermi distribution \cite{density_fermitype}. 
\begin{equation}
\rho (r) = \rho_0 [\frac{1+w(r/R)^{2}}{1+exp[(r-R)/a]}]\,.\\
\label{eqn1}
\end{equation}
Here $\rho_0$ is the nuclear matter density in the centre of the nucleus, R is  the radius of the nucleus from its centre. The parameter, a, is the skin depth or surface thickness, r is a position parameter and distance of any point from centre of the nucleus, 
and w is the deviation from a smooth spherical surface.
$\rm{Au}^{197}$ or $\rm{Pb}^{208}$ nucleus is assumed here to have uniform distribution of nucleons in its approximately spherical volume and smooth surface, so that $\displaystyle w$ can be taken to be zero.
This reduces~eqn.\ref{eqn1} to Woods-Saxon \cite{wdsx} distribution, which has been used in HIJING in most cases. This may be written as:\\
\begin{equation}
\rho (r) = \frac{\rho_0}{1+exp[(r-R)/a]}\,.\\
\label{eqn2}
\end{equation}


When we use an axially symmetric or prolate deformed nucleus (viz. $\rm U^{238}$, $\rm Xe^{56}$ etc.), nuclear radius $R$, has been modified to include spherical harmonics. The modified Woods-Saxon nuclear radius \cite{wdsx_deform} may be written as:
\begin{equation}
R_{A\Theta} = R[1+ \beta_2 Y_{20}(\theta)+\beta_4 Y_{40}(\theta) ],
\label{eqn3}
\end{equation}
where the symbols $\beta_i$ are deformation parameters. In case of Xenon nucleus, we have used deformation parameters, $\displaystyle\beta_2$= 0.162 and $\displaystyle\beta_4$= -0.003 from Ref. \cite{atomic_data_table}.
The spherical harmonics, $Y_{20}$, and $Y_{40}$ are given by~\cite{speherical_harmonic}, \\
\begin{eqnarray}
Y_{20}(\theta) &=& \frac{1}{4} \sqrt{\frac{5}{\pi}}(3 \ cos^2\theta -1)\nonumber\\
Y_{40}(\theta) &=& \frac{3}{16\sqrt\pi}(35 \ cos^4\theta -30 \ cos^2\theta+3)\,.
 \end{eqnarray}

The positions of nucleons within the distribution, $\rho(r)$, are sampled using the volume element 
$r^2 sin \theta \ dr \ d\theta \ d\phi$ \cite{thesis_chris, ptribedy}. For random orientation of nuclei, position configurations are sampled with both polar angle, (angle between major axis and beam axis) $\Theta$ in $[0, \pi]$) and azimuthal angle, (angle between major axis and impact parameter) $\Phi$ within limits $[0, 2\pi]$. Both target and projectile nuclei are rotated event-by-event in azimuth and polar space. 
In this paper, calculations have been done only with random orientation which means, unpolarized and averaged value over random $\Theta$ and $\Phi$~\cite{define_config}. In Fig. \ref{fig0}, we show normalized nuclear density profile of Xenon with and without deformations. The lower panel shows the ratio of the density profile with nuclear deformation, with respect to the case of no-deformation$^{[\ref{myfootnote}]}$.

\footnotetext[1]{\label{myfootnote} Initial eccentricity for Xe+Xe, (0-5)\% central collisions, $\epsilon_2$ $\approx$ 0.24$\pm$0.03 in random $\theta$ and $\phi$ orientations.}


The initial low-momentum partons which are separated from high momenta partons by a momentum cut-off of $p_{\rm T}$ = 2.0 GeV/c, are produced from parametrized coloured string fragmentation mechanisms. The high-$p_{\rm T}$ particles (usually minijets) are sensitive to this momentum cut-off~\cite{spmb}. The produced particles are then initiated into parton transport part, ZPC (Zhang's Parton Cascade Model)~\cite{amptzpc}, which transports the quarks and gluons using Boltzmann transport equation which is given by,
\begin{eqnarray}
p^{\mu}\partial_{\mu}f(x,p,t)=C[f]
\end{eqnarray}
The leading order equation showing interactions among partons is approximately given by,
\begin{equation}
\frac{d\hat{\sigma}_{gg}}{d\hat{t}}\approx \frac{9\pi\alpha_s^2}{2(\hat{t}-\mu^2)^2}\,.
\end{equation}
Here $\sigma_{gg}$ is the gluon scattering cross-section, $\alpha_s$ is the strong coupling constant used in above equation, and $\mu^2$ is the cutoff used to avoid infrared divergences which can occur if the momentum transfer, $\hat{t}$, goes to zero during scattering. In the String Melting version of AMPT (AMPT-SM), melting of coloured strings into low momentum partons also take place at the start of the ZPC and are calculated using Lund FRITIOF model of HIJING. This melting phenomenon depends upon spin and flavour of the excited strings. The resulting partons undergo multiple scatterings which take place when any two partons are within distance of minimum approach which is given by $\displaystyle d\,\leq\,\sqrt{\sigma/\pi}$, where $\sigma$ is the scattering cross-section of the partons. In AMPT-SM, the transported partons are finally hadronized using coalescence mechanism~\cite{amptreco}, when two (or three) quarks sharing a close phase-space combine to form a meson (or a baryon). 
The coalescence in AMPT can be shown by the following equation (for e.g. meson), 
\begin{eqnarray}
\frac{d^3N}{d^3p_M}=g_M\int{d^3x_1d^3x_2d^3p_1d^3p_2\,f_q(\vec{x}_1,\vec{p}_1)}f_{\bar{q}}(\vec{x}_2,\vec{p}_2)\nonumber\\
                            \delta^3(\vec{p}_M-\vec{p}_1-\vec{p}_2)\,f_M(\vec{x}_1-\vec{x}_2,\vec{p}_1-\vec{p}_2).
\end{eqnarray}
\begin{figure*}[ht]
\includegraphics[height=15em]{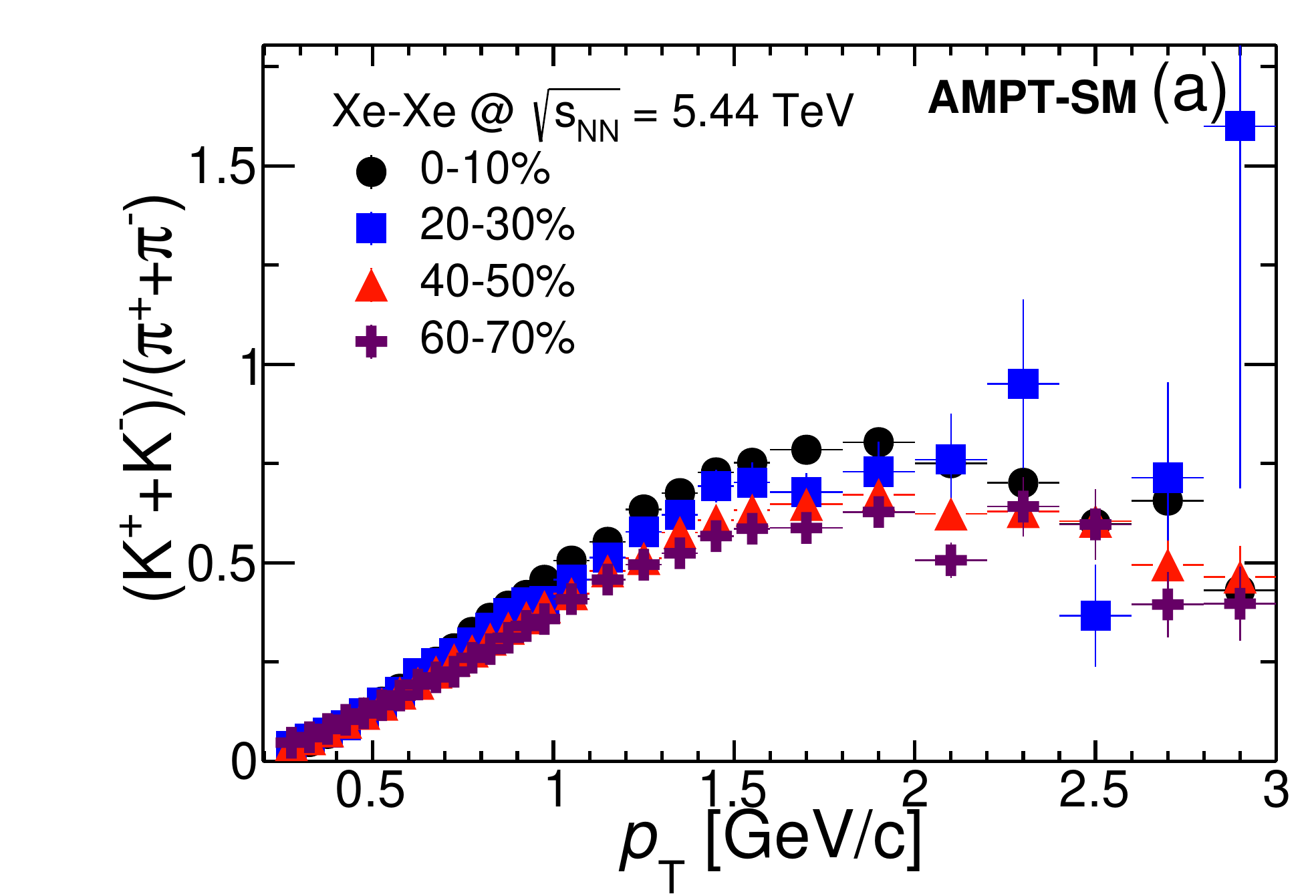}
\includegraphics[height=15em]{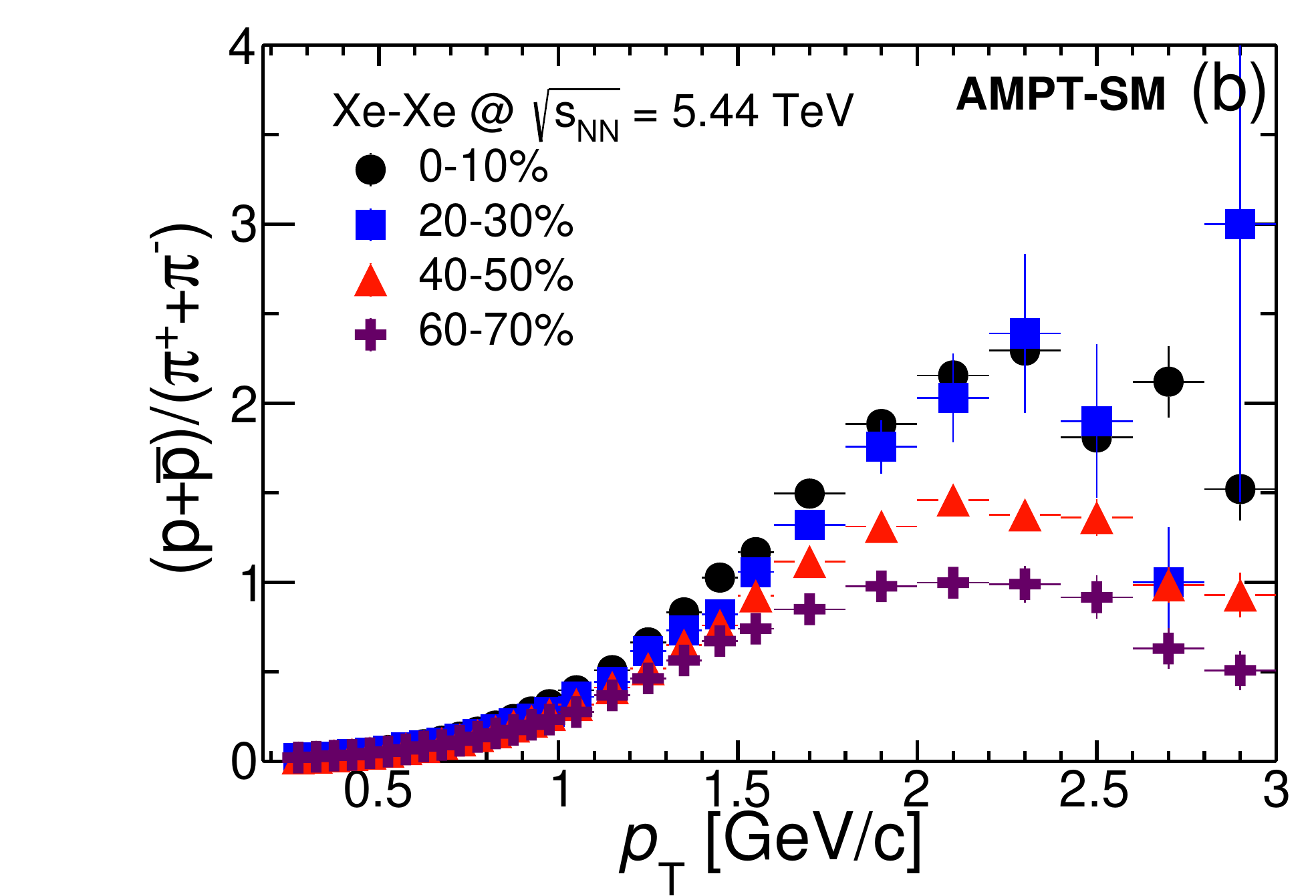}
\includegraphics[height=15em]{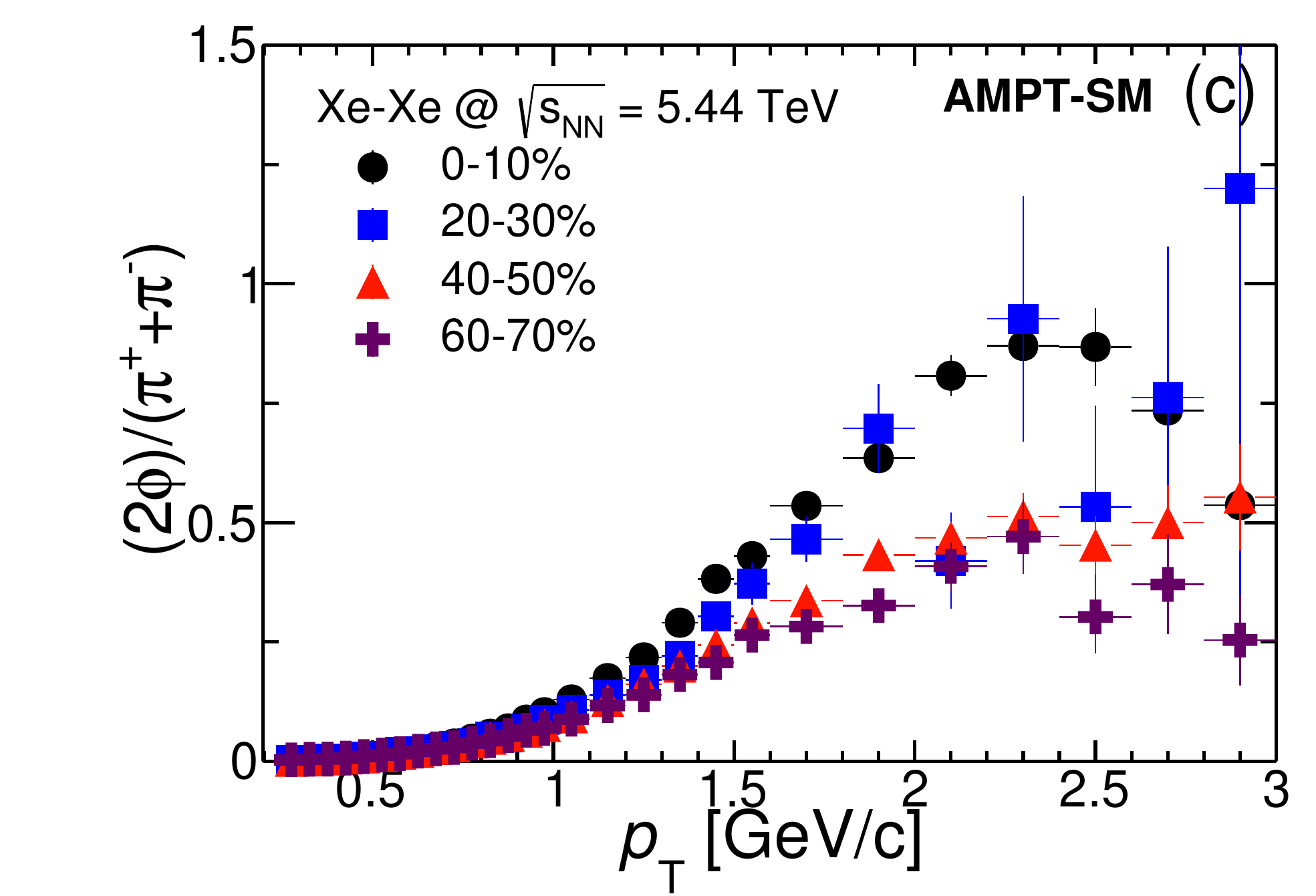}
\includegraphics[height=15em]{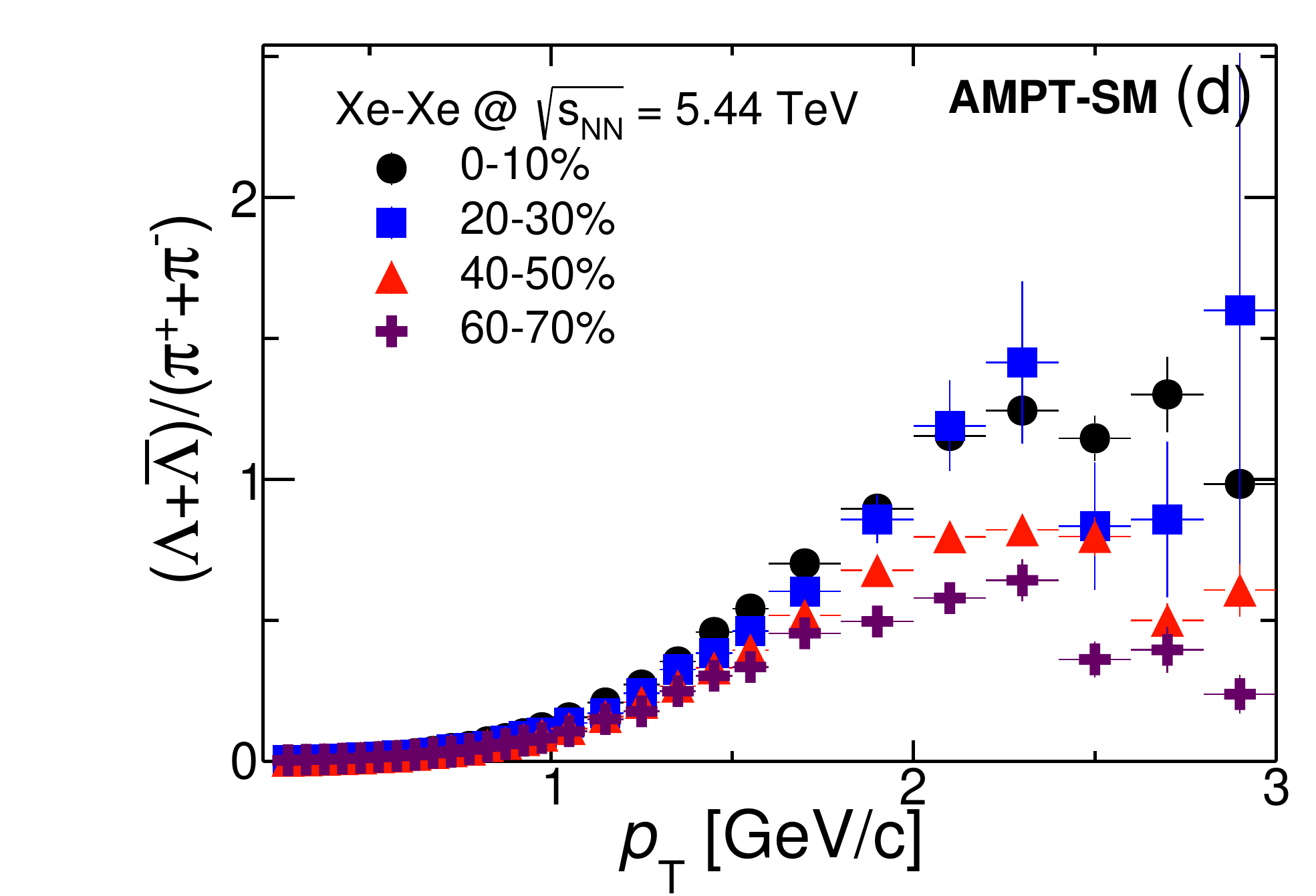}
\caption[]{(Color online) $p\rm{_T}$-differential particle ratios of kaons (a), protons (b), $\phi$ (c) and $\Lambda$ (d) to pions in Xe+Xe collisions at $\sqrt{s_{\rm NN}}$ = 5.44 TeV. Different symbols show various centrality bins. The vertical lines in the data points are the statistical uncertainties.}
\label{fig3}
\end{figure*}
Here $g_M$ is the meson degeneracy factor, $f_q$'s are the quark distributions after the evolution, and $f_M$ is the coalescing function commonly called Wigner functions~\cite{amptreco}.
The produced hadrons further undergo evolution in ART mechanism~\cite{amptart1, amptart2} via meson-meson, meson-baryon and baryon-baryon interactions, before final spectra can be observed. The default version of AMPT known as AMPT-Def, where instead of coalescing the partons, we have fragmentation mechanism using Lund fragmentation parameters $a$ and $b$ used for hadronizing the transported partons. However, it can be shown that particle flow and spectra at the mid-$p_T$ regions are well explained by quark coalescence mechanism for hadronization~\cite{ampthadron1,ampthadron2,ampthadron3}. We have used AMPT-SM mode for our calculations. We will return to the discussion of our choice in results and discussion section. We have used the AMPT version 2.26t7 (released: 28/10/2016) in our current work.
It is worthwhile to mention that earlier studies of particle elliptic flow in Pb+Pb collisions with AMPT showed greater match with experimental data when large partonic scattering cross-section ($\sigma_{gg}$ $\approx$ 10 mb) is taken~\cite{Feng:2016emh,Tripathy:2018bib}. As expected, results with $\sigma_{gg}$ $\approx$ 10 mb shows greater $v_2$ than 3 mb. While, taking rapidity, $\eta$, as the variable, the difference in 10 mb and 3 mb results can be seen as a constant multiplication factor, particularly in the central rapidity region \cite{Tripathy:2018bib}. In the present work, we have fixed $\sigma_{gg}$ = 10 mb as  cross-section for our calculations and the estimation of identified particle ratios. The Lund string fragmentation parameters $a$ and $b$ are kept fixed at their default values of 2.2 and 0.5/GeV$^2$, respectively.  It should be noted here that as we intend to study $\phi$ and $\rm{K}_{s}^0$ in the final state, we have kept hadron level decay flagged as off  for $\phi$ and $\rm{K}_{s}^0$ throughout our analysis. As expected, this flag affects the total particle multiplicity, when studied as a function of collision centrality. The $N_{\rm part}$-normalized integrated yield (dN/dy) as a function of $N_{\rm part}$ (centrality) seems to follow a monotonic decrease with an increase of $N_{\rm part}$ for pions, kaons and protons. However, this seems to be almost independent of centrality for $\phi$ and $\Lambda$. Furthermore, we have checked explicitly that when the decay of $\phi$ and $\rm{K}_{s}^0$ is allowed the dN/dy for all the identified particles seems to show a monotonic rise with collision centrality. 

\section{Results and discussions}
As described in the previous section, we have generated events using AMPT model in different centralities for Xe+Xe collisions at the mid-rapidity for $\sqrt{s_{\rm NN}}$ = 5.44 TeV, so that the results could be compared with the corresponding ALICE/CMS experimental data, when become available. We study the $p\rm{_T}$-spectra and integrated yield of identified particle production such as ($\pi^{+}+\pi^{-}$), (K$^{+}$+K$^{-}$), $\rm{K}_{s}^0$, (p+$\bar{\rm{p}}$), $\phi$ and ($\Lambda + \bar{\Lambda}$). We also study the $p\rm{_T}$-differential and $p\rm{_T}$-integrated particle ratios to ($\pi^{+}+\pi^{-}$) and (K$^{+}$+K$^{-}$). From here onwards, ($\pi^{+}+\pi^{-}$), (K$^{+}$+K$^{-}$), (p+$\bar{\rm{p}}$) and ($\Lambda + \bar{\Lambda}$) are denoted as pions ($\pi$), kaons (K), protons (p) and $\Lambda$, respectively. As the particle production mechanisms are highly dependent on the transverse momentum range, {\it e.g.}, when at intermediate-$p_T$, coalescence becomes the major mechanism, at high-$p_T$, the fragmentation takes over, it is worth studying $p_T$-differential particle ratios. This is the prime focus of the present work. 


%
In Fig.\ref{fig1}, we have shown $p\rm{_T}$-spectra of identified hadrons for 0-10\% central collisions of Xe+Xe at mid-rapidity ($|\eta|\rm{<}$0.8). Different symbols represent the $p\rm{_T}$ spectra for various particle species. Pions, being the lightest hadron, the production is maximum. At low-$p\rm{_T}$, we observe a mass-dependent behavior of the produced particles. The global mass ordering is violated as the production of $\phi$ is lesser compared to $\Lambda$. This behavior is similar to the experimental data from ALICE at the LHC \cite{ALICEfig2}. While pions show almost an exponentially decreasing behavior, other particles' spectra show dip at $p_T <$ 0.5 GeV/c and they approach the pion spectra at intermediate $p\rm{_T}$. This behavior could be due to the radial flow effects in a medium as the radial flow pushes the particles from low-$p\rm{_T}$ to intermediate-$p\rm{_T}$ \cite{radialflow1}.
 Also, these shapes of the $p_T$-spectra may be due to the coalescence mechanism \cite{coalescence1} at the low and intermediate momenta and/or the reason might also be the production of high-$p\rm{_T}$ jets \cite{ampthadron1, recojet} caused by fragmentation mechanism but it's effects are mostly found beyond intermediate momentum region. 
It would be interesting to study the kinetic freeze-out properties in Xe+Xe collisions. Taking $(0-10)\%$ centrality class and pion $p_{\rm T}$-spectra, we observe the average radial flow velocity to be $<\beta_r> = 0.45 \pm 0.04$ and the $T_{\rm kin} = 109 \pm 12$ MeV. This estimation is done by fitting Boltzmann-Gibbs Blast Wave model (BGBW) \cite {BGBW} to the $p_{\rm T}$-spectra up to $p_{\rm T} \sim$ 3 GeV/c. We have assumed a linear velocity profile in BGBW, which considers the produced fireball as a hard sphere uniform-density particle source. A centrality dependent study shows that both $T_{\rm kin}$ and $<\beta_r>$ are centrality dependent- higher radial flow in central collisions resulting in drop in $T_{\rm kin}$, as was earlier observed in heavy-ion collisions \cite{Abelev:2008ab}.

%


Figure ~\ref{fig3} shows $p\rm{_T}$-differential particle ratios of kaons, protons, $\phi$ and $\Lambda$ to pions at different centralities. All the particle ratios with respect to $\pi$ increases as a function of $p\rm{_T}$. Considering K to $\pi$ ratio as a measure of strangeness, we observe enhancement of strangeness production as a function of  $p\rm{_T}$. This enhancement has a weak-dependence on centrality at low-$p\rm{_T}$, while it strongly depends on centrality at intermediate-$p\rm{_T}$ region. At intermediate-$p\rm{_T}$, the strangeness production is maximum for the central collisions and it decreases with the centrality. A similar behavior is observed for the case of $\Lambda$ to $\pi$ ratio, where $\Lambda$ has the same strangeness content as of K. However, K to $\pi$ ratio rises more rapidly than $\phi$ to $\pi$ ratio which is more gradual. The reason may be due to the higher probability for a strange quark to find a up or a down quark to form kaons rather than find its anti-strange quark to form $\phi$ meson at low momentum region. As we move from low particle momenta to intermediate momentum region when particle momentum is comparable to or more than the mass of strange ($s$) quark, the probability of $s\bar{s}$ production increases considerably so that $\phi$ to $\pi$ ratio is found to increase. However at higher momentum, more number of $u$ and $d$ quarks are also produced as  compared to $s$ quark so that both ratios also start to drop beyond $p_T\approx$ 2 GeV. Similar trend of particle ratios are also observed in p-Pb and Pb-Pb collisions~\cite{ppALICE1,pPbALICE1}. 

Figure ~\ref{fig3}(b) shows the ratio of p to $\pi$, which is a ratio between lightest baryons to lightest mesons, which serves as a proxy of baryon to meson ratio. We have found that the trend is similar as other ratios but the values are quite different for p to $\pi$ ratios. For most central Xe+Xe collisions, the p to $\pi$ and $\Lambda$ to $\pi$ ratios are more than 1 in the intermediate-$p\rm{_T}$ region, which indicates that the baryon production is more compared to lightest meson in the intermediate-$p\rm{_T}$ region. We will revisit about this behavior at the end of this section.

\begin{figure}[ht]
\includegraphics[height=15em]{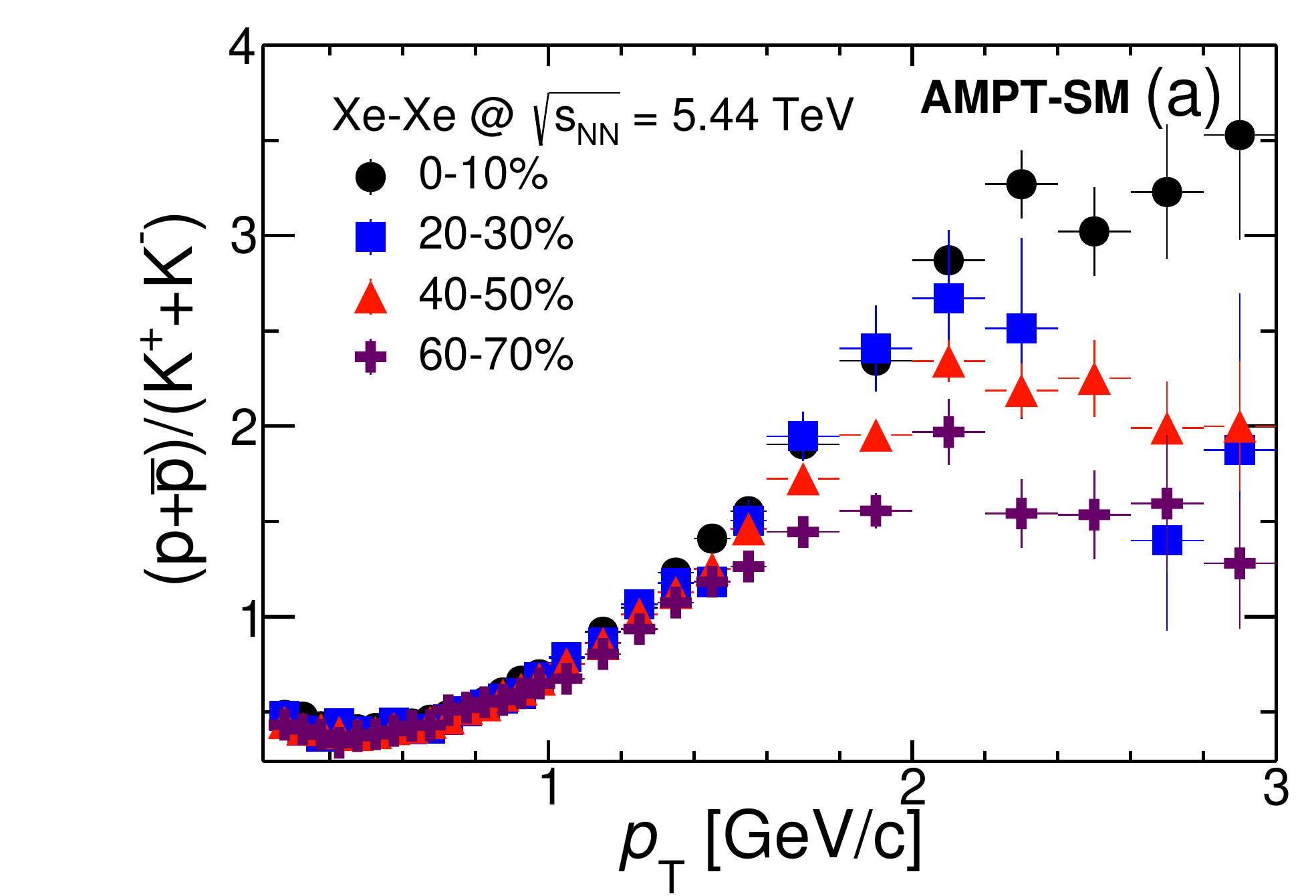}
\includegraphics[height=15em]{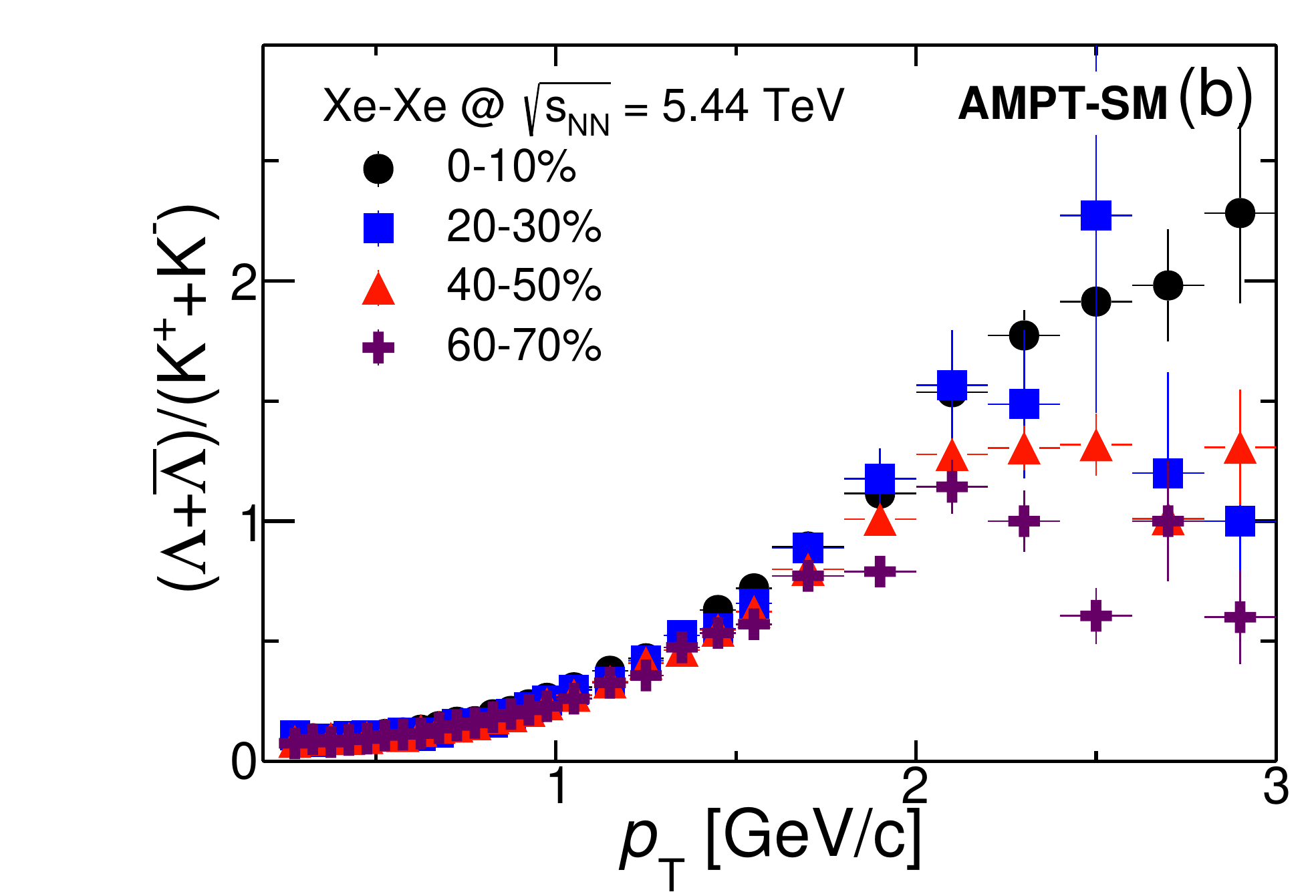}
\caption[]{(Color online) $p\rm{_T}$-differential p to K (a) and $\Lambda$ to K (b) ratio for various centrality bins in Xe+Xe collisions at $\sqrt{s_{\rm NN}}$ = 5.44 TeV. The vertical lines in the data points are the statistical uncertainties.}
\label{fig4}
\end{figure}
Figure ~\ref{fig4} represents the ratios of baryons over lightest strange meson, K. The upper panel of the figure shows p to K and the lower panel of the figure shows $\Lambda$ to K ratios as a function of $p\rm{_T}$ for collisions at different centralities. Both the ratios are independent of centrality at low-$p\rm{_T}$ while they depend on centrality in the intermediate-$p\rm{_T}$ ranges. For a given $p\rm{_T}$-bin, after $p\rm{_T}>$~1 GeV/c, the ratios decrease with centrality. This trend is similar to the particle ratios with respect to $\pi$ in Fig.~\ref{fig3}.

\begin{figure}[ht]
\includegraphics[height=15.5em]{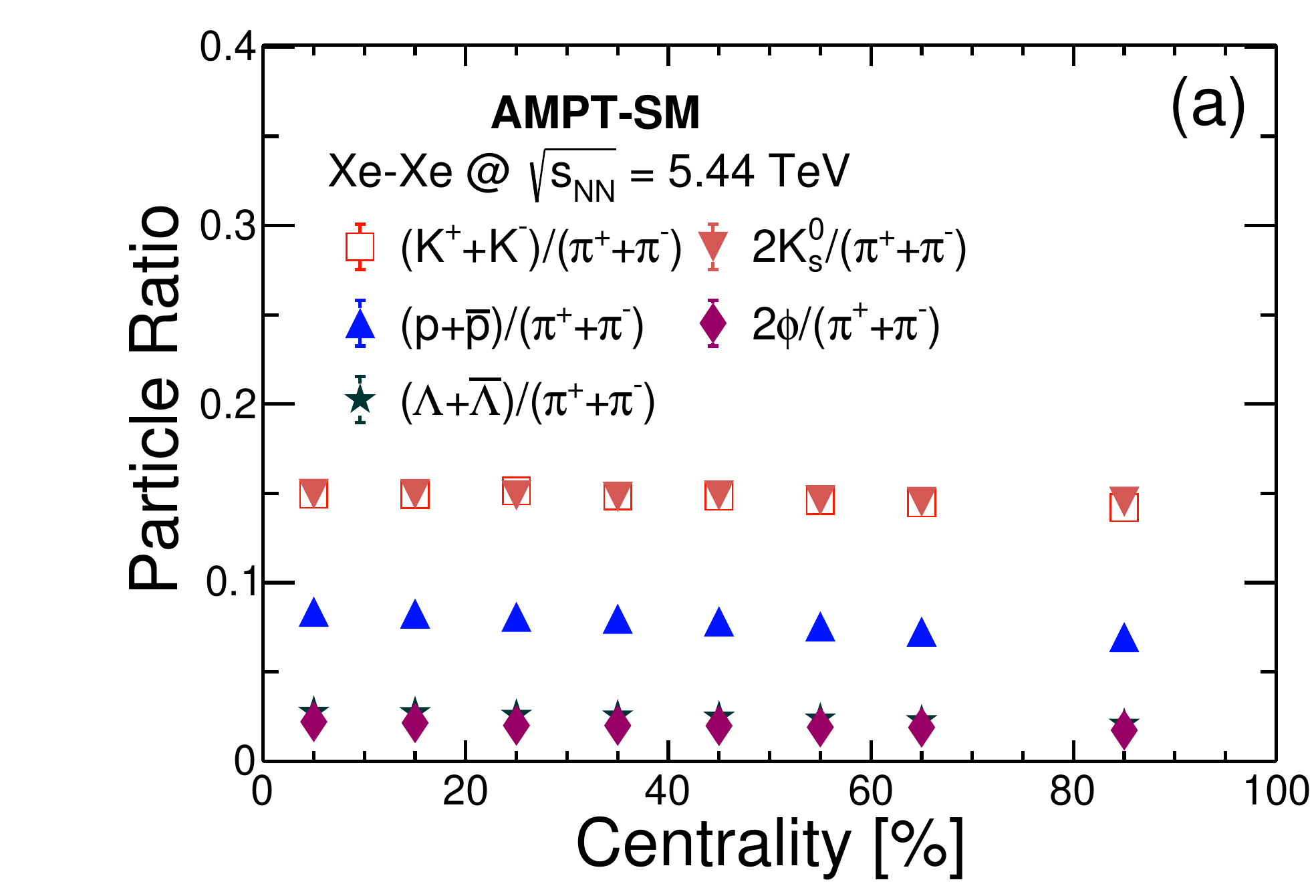}
\includegraphics[height=15.5em]{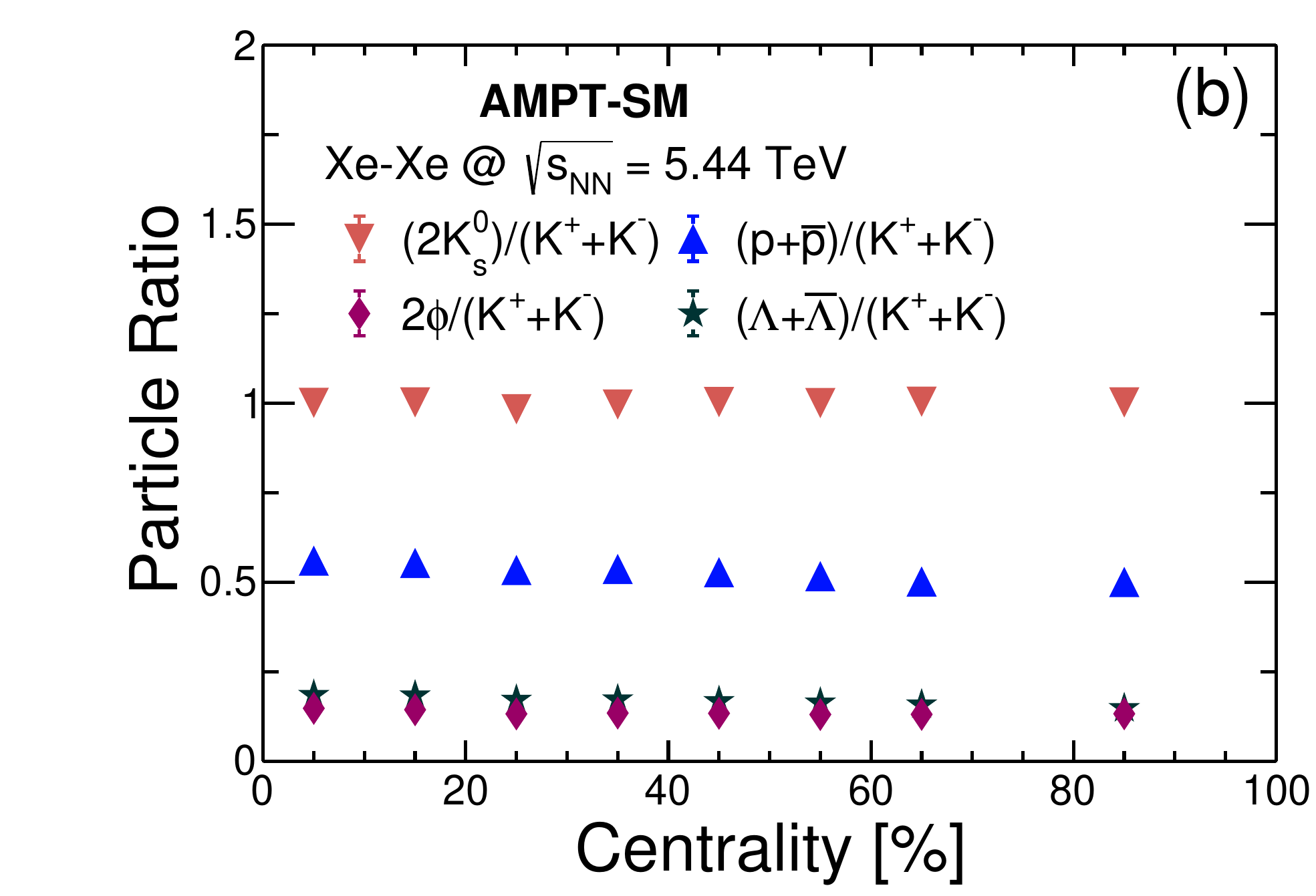}
\caption[]{(Color online) $p\rm{_T}$-integrated ratios of identified particles to pions (a) and kaons (b) as a function of centrality. Different symbols are for different particles. The statistical uncertainties are within symbol sizes.}
\label{fig5}
\end{figure}

Figure~\ref{fig5} shows the $p\rm{_T}$-integrated ratios of identified hadrons over pions and kaons as a function of centrality. It is very interesting to see that while differential particle ratios show strong dependence with centrality (for $p\rm{_T} > 1$ GeV/c), the integrated particle ratios show no centrality dependence. This indicates that the relative particle production with respect to pion does not depend on the centrality. This is due to the fact that the integrated yield is dominated by low-$p\rm{_T}$ ($p\rm{_T} <$ 1 GeV/c) particles. Assuming both centrality and charged-particle multiplicities are used as a proxy for the system size, the centrality-dependent particle ratios of p to $\pi$ and $\phi$ to $\pi$ as a function of centrality in Xe+Xe collisions at $\sqrt{s_{\rm NN}}$ = 5.44 TeV reproduces qualitatively (within uncertainties) the preliminary results as a function of charged-particle multiplicity of ALICE at the LHC~\cite{Bellini:2018khg,Tripathy:2018ehz}. Also, the trend of these ratios are similar to the experimental data in Pb-Pb collisions at $\sqrt{s_{\rm NN}}$ = 2.76 and 5.02 TeV reported by ALICE~\cite{Abelev:2013vea, Bellini:2018khg,Tripathy:2018ehz, Dash:2018cjh, Abelev:2014laa}.

\begin{figure}[ht]
\includegraphics[height=17.5em]{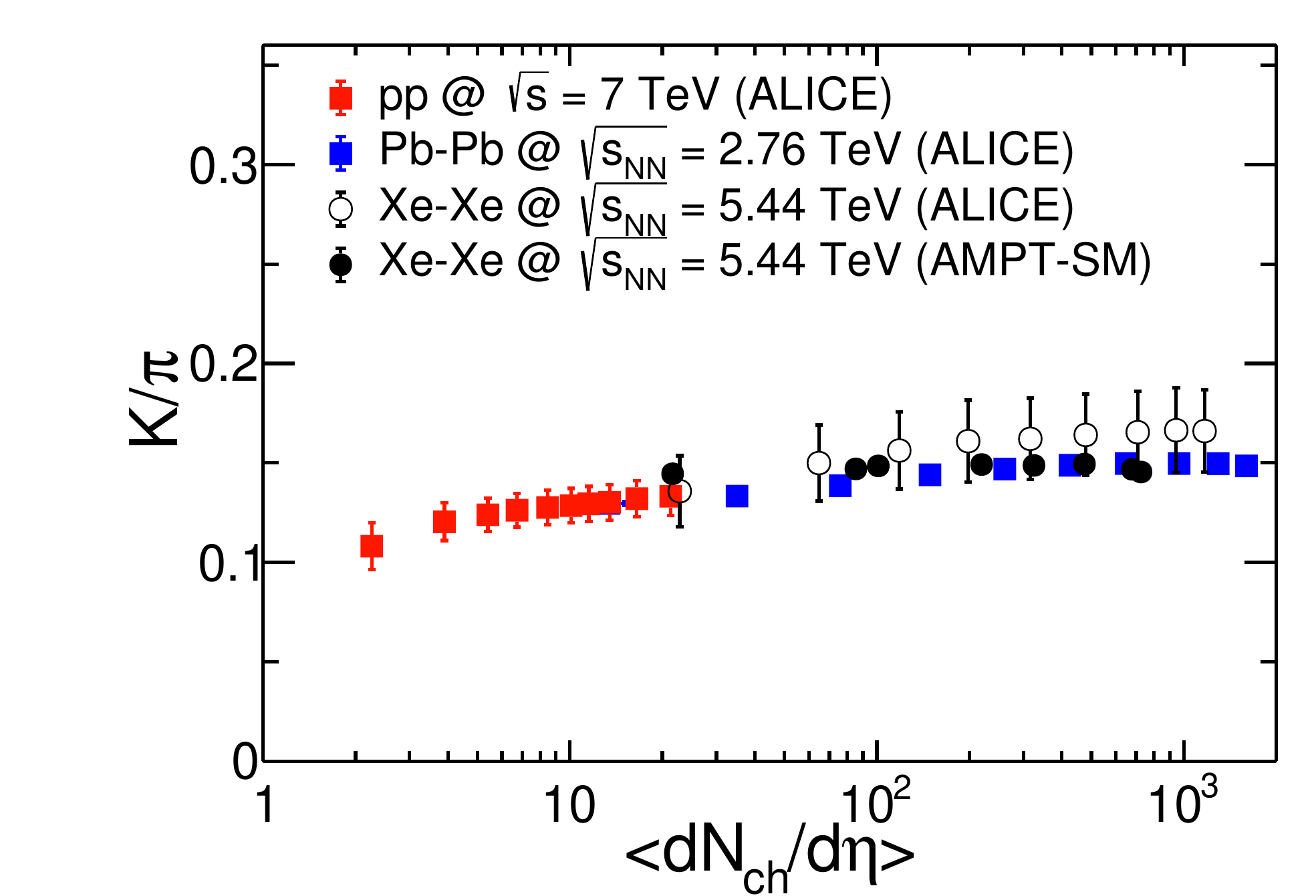}
\caption[]{(Color online) $p\rm{_T}$-integrated Kaon to pion ratio as a function of charged-particle multiplicity for pp collisions at $\sqrt{s}$ = 7 TeV, Pb+Pb collisions at $\sqrt{s_{\rm NN}}$ = 2.76 TeV and Xe+Xe collisions at $\sqrt{s_{\rm NN}}$ = 5.44 TeV. The ratios for p+p, Xe+Xe and Pb+Pb collisions are from ALICE experimental data~\cite{ALICE:2017jyt,Bellini,Abelev:2013vea}. The ratio for Xe+Xe collisions from AMPT-SM seems to agree with the experimental data within uncertainties. The experimental data from p+p collisions are for $\rm K_{s}^{0}/\pi$ while for others, the ratio is (K$^{+}$+K$^{-}$)/($\pi^{+}+\pi^{-}$).}
\label{fig5.1}
\end{figure}

Figure~\ref{fig5.1} shows the $p\rm{_T}$-integrated Kaon to pion ratio as a function of charged-particle multiplicity for p+p collisions at $\sqrt{s}$ = 7 TeV, Pb+Pb collisions at $\sqrt{s_{\rm NN}}$ = 2.76 TeV and Xe+Xe collisions at $\sqrt{s_{\rm NN}}$ = 5.44 TeV. The ratios for p+p and Pb+Pb collisions are from experimental data~\cite{ALICE:2017jyt,Abelev:2013vea}. The ratio for Xe+Xe collisions are from AMPT-SM and are compared to the ALICE experimental results \cite{Bellini}. We observe that AMPT-SM predictions seem to match with the experimental data within uncertainties. Although data are from different energies and  collision systems ($viz.$ Pb+Pb, Xe+Xe and p+p), the proxy of strangeness enhancement, K/$\pi$ seems to follow a scaling with final state charged particle multiplicity, which in our opinion, is a very good observation.  On a finer scale, the K to $\pi$ ratio seems to show an increasing trend, which indicates strangeness enhancement with system size. It would be interesting to have experimental data for p+p collisions at $\sqrt{s}$ = 5.02 TeV and p+Pb collisions at $\sqrt{s_{\rm NN}}$ = 5.02 TeV to have a proper conclusion of this interesting observation.

\begin{figure}[ht]
\includegraphics[height=15em]{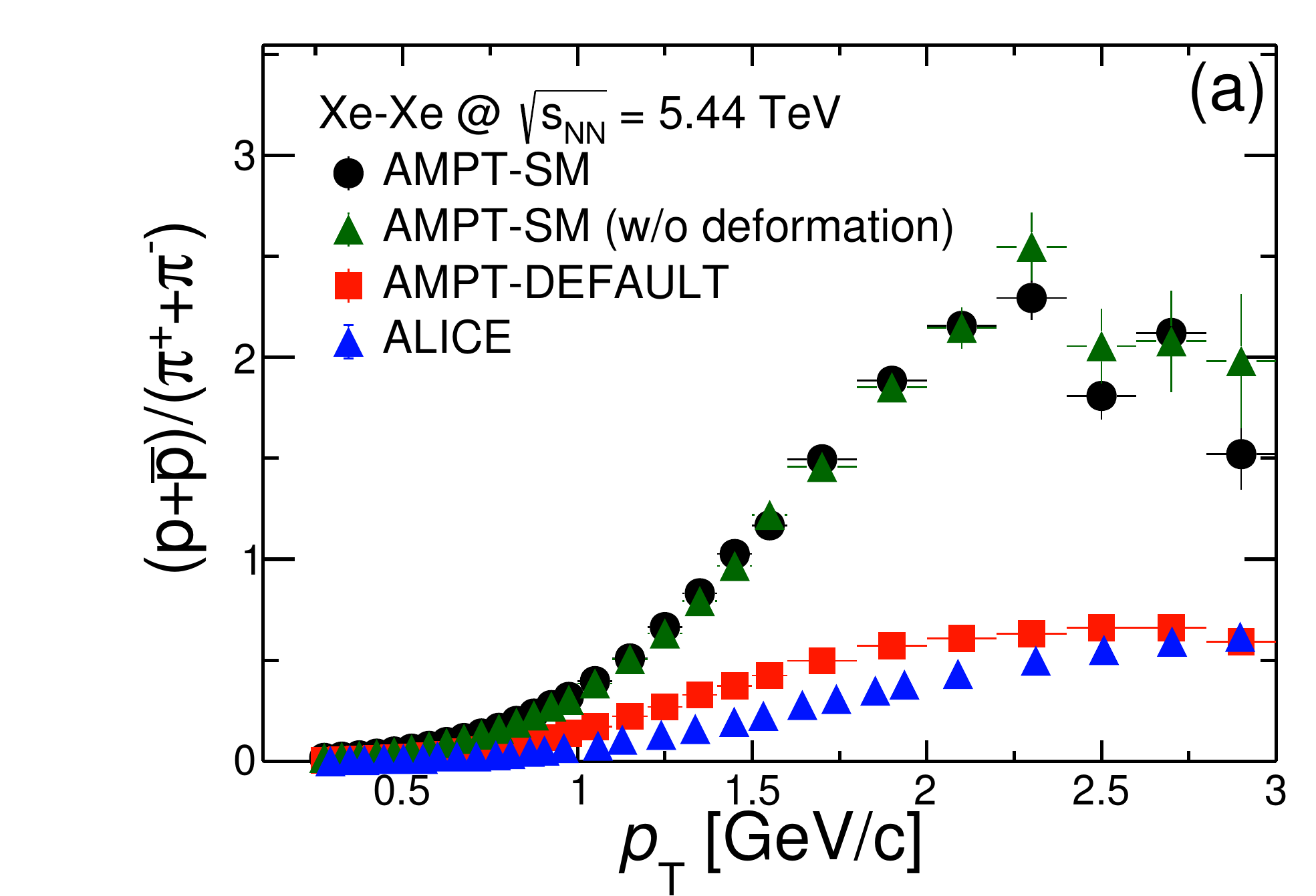}
\includegraphics[height=15em]{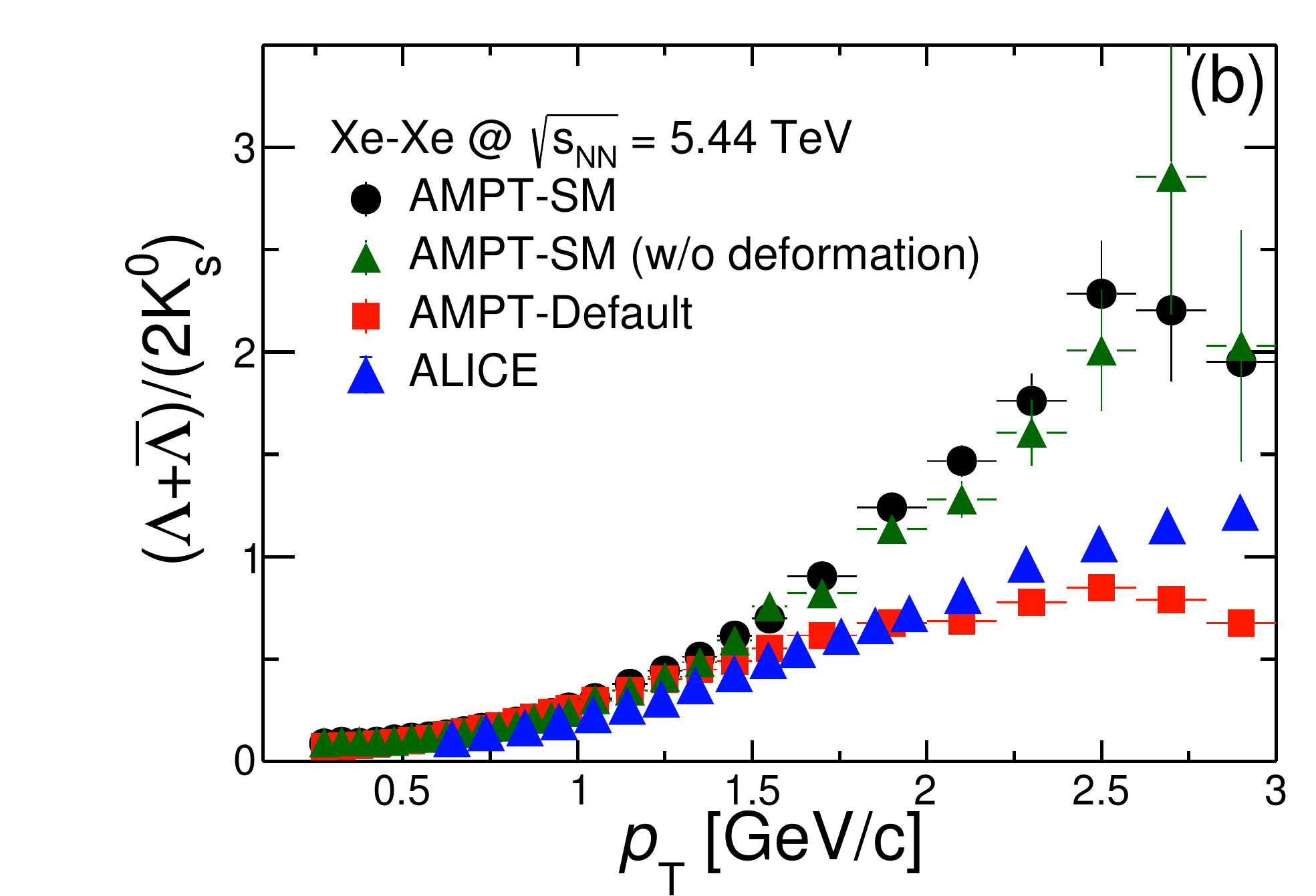}
\includegraphics[height=15em]{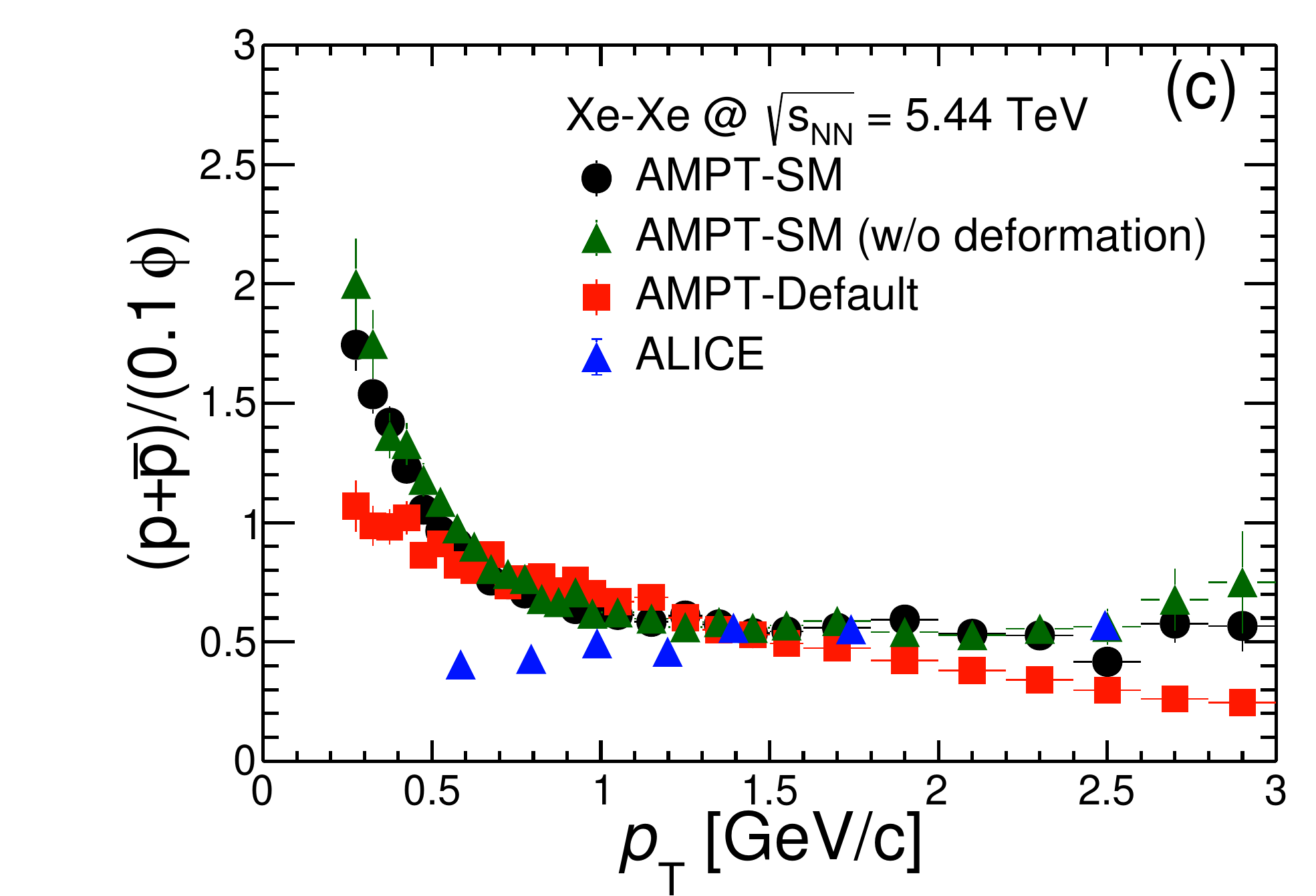}
\caption[]{(Color online) $p\rm{_T}$-differential p to $\pi$ (a), $\Lambda$ to $\rm{K}_{s}^0$ (b) and p to $\phi$ (c) ratios for most central (0-10\%) Xe+Xe collisions at $\sqrt{s_{\rm NN}}$ = 5.44 TeV. Red and black markers are predictions from AMPT-Default and AMPT-SM, respectively. The ALICE preliminary data~\cite{Bellini:2018khg} are shown in blue markers. The error bars in the results from models are the statistical uncertainties.}
\label{fig6}
\end{figure}
Figure~\ref{fig6} shows the comparison of $p\rm{_T}$-differential p to $\pi$, $\Lambda$ to $\rm{K}_{s}^0$ and p to $\phi$ ratios for most central (0-10\%) Xe+Xe collisions from AMPT-SM with the AMPT-Default version. Also, they are compared with the preliminary experimental data~\cite{Bellini:2018khg}. For p to $\pi$ and $\Lambda$ to $\rm{K}_{s}^0$ ratios, it is observed that the AMPT-Default version is more closer to the experimental data than that of AMPT-SM specially for $p_{T} >$ 1 GeV/c. It seems that although AMPT-SM describes the elliptic-flow of the charged particles of experimental data \cite{Lin:2014uwa}  better than the AMPT-Default but in case of stable particle ratios, AMPT-Default does a better job than the AMPT-SM. This may be due to the coalescence mechanism involved in AMPT-SM, which affects the particle production at intermediate-$p\rm{_T}$. However, in the case of p to $\phi$ ratio both the versions of AMPT fails to explain the experimental data at low-$p\rm{_T}$. At intermediate and high-$p\rm{_T}$, AMPT-SM prediction is closer to the experimental data. According to hydrodynamics-inspired models, particles with similar masses should have similar particle spectra at low-$p\rm{_T}$.  It is found that the ratio is flat for experimental data over all the $p_{T}$ region, whereas for AMPT it decreases upto $p_{T} \sim$ 1 GeV/c and then remain flat over higher $p_{T}$ region. 

%


We have explicitly observed that the particle ratios are independent of nuclear deformation in Xe+Xe collisions. However, it should also be mentioned here that the identified particle $p_{\rm{T}}$-spectra might be sensitive to nuclear deformation for central Xe+Xe collisions. For the case of deformation, the particle yield ratios are found to be comparable to the case of a spherical Xe nucleus. These findings indicate that nuclear deformation is insensitive to chemical freeze-out in Xe+Xe collisions. Similar results are observed when particle ratios calculated from U+U collisions are compared to Au+Au collision systems~\cite{Tripathy:2018rln}.

\begin{figure}[ht]
\includegraphics[height=18em]{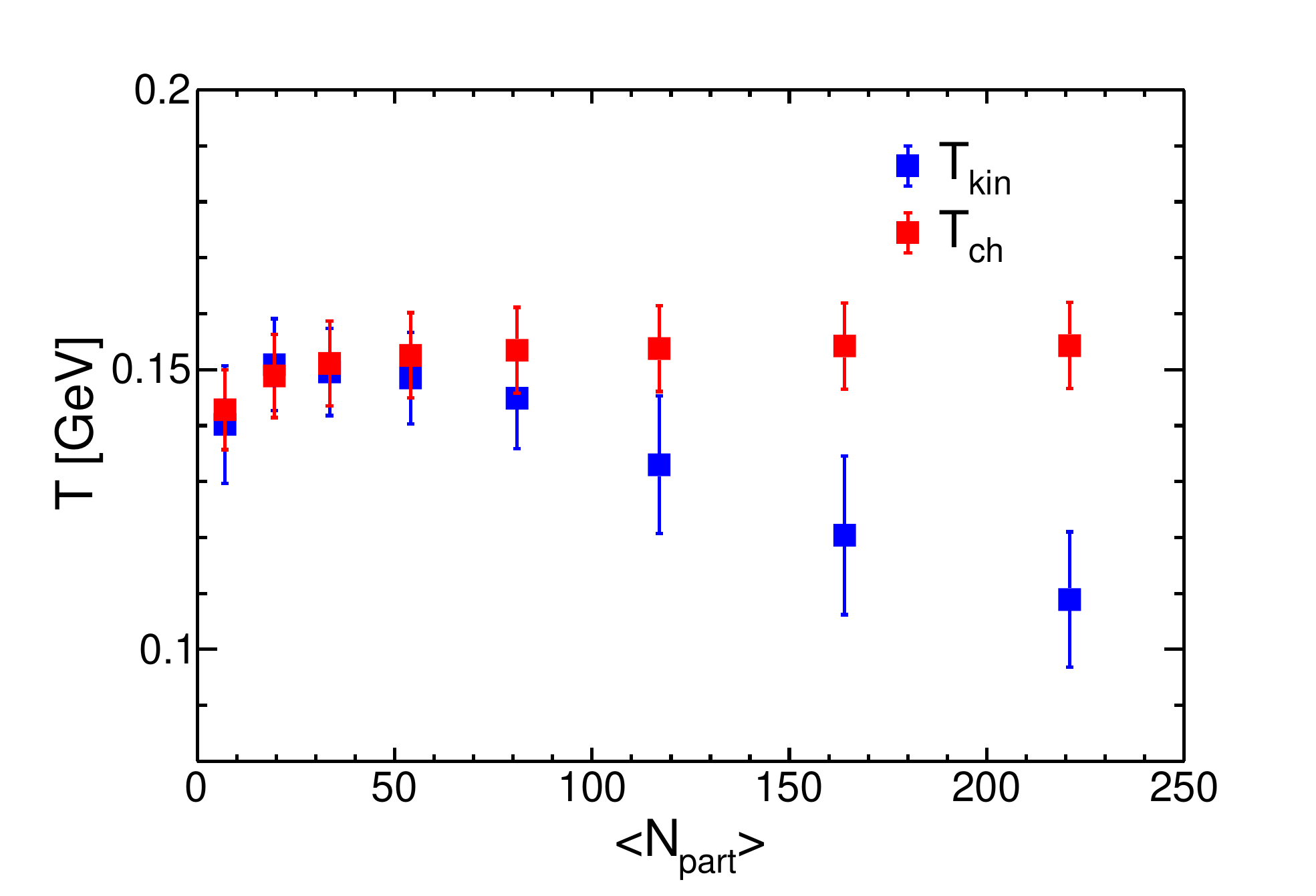}

\caption[]{(Color online) The kinetic freeze-out temperature, $T_{\rm kin}$ and the chemical freeze-out temperature, $T_{\rm ch}$ as a function of collision centrality in Xe+Xe collisions at $\sqrt{s_{\rm NN}}$ = 5.44 TeV.}
\label{fig8}
\end{figure}

We study chemical freeze-out temperature, $T_{\rm ch}$ as a function of collision centrality measured through $<N_{\rm part}>$ in Xe+Xe collisions at $\sqrt{s_{\rm NN}}$ = 5.44 TeV using AMPT. This is shown in Fig. \ref{fig8} along with a comparison of kinetic freeze-out temperature, $T_{\rm kin}$. The $T_{\rm ch}$ for $(0-10)\%$ centrality is found to be around $154 \pm 8$ MeV, which is comparable with p+Pb collisions at $\sqrt{s_{\rm NN}}$ = 5.02 TeV \cite{Sharma:2018jqf}. As expected, the present analysis of particle ratios reveal $T_{\rm ch}$ to be independent of collision centrality in Xe+Xe collisions. Here we have taken the discussed particle ratios and have assumed a grand canonical ensemble in the thermal model \cite{Wheaton:2011rw}  taking $\mu_B =0$ and keeping  $T_{\rm ch}$, the strangeness saturation factor, $\gamma_s$ and the fireball radius as the free parameters. The kinetic freeze-out temperature is found to be highly dependent on collision centrality.

\section{Summary}
We have studied the $p\rm{_T}$-spectra, integrated yield and particle ratios of identified particles for Xe+Xe collisions at $\sqrt{s_{\rm NN}}$ = 5.44 TeV using AMPT. Our findings are the following:
\begin{enumerate}
\item We have reported the simulation studies of identified particle production in Xe+Xe collisions at $\sqrt{s_{\rm NN}}$ = 5.44 TeV using AMPT model. This can be compared with experimental data, when become available. In particular, the effect of nuclear deformation warrants a simulation study, which through this work, would give a better understanding to experimental findings.

\item A BGBW analysis of pion $p_{\rm T}$-spectra up to $p_{\rm T} \sim$ 3 GeV/c for $(0-10)\%$ centrality class shows the radial flow velocity, $<\beta_r>  = 0.45 \pm 0.04$ and the kinetic freeze-out temperature, $T_{\rm kin} = 109 \pm 12$ MeV. As observed earlier in heavy-ion collisions at RHIC, the radial flow velocity decreases  and the the kinetic freeze-out temperature increases towards peripheral collisions.

\item We observe enhancement of strangeness production as a function of $p\rm{_T}$. This enhancement has a weak-dependence on centrality at low-$p\rm{_T}$, while it strongly depends on centrality at intermediate-$p\rm{_T}$ region.

\item We observe that the differential particle ratios show strong dependence with centrality (for $p\rm{_T} > 1$ GeV/c) while the integrated particle ratios show no centrality dependence.

\item It is indeed interesting to note that the proxy of strangeness enhancement, K/$\pi$ ratio, when studied as a function of final state charged particle multiplicity for p+p, Xe+Xe and Pb+Pb collisions at different collision energies at the LHC, shows a scaling behavior indicating that the final state multiplicity drives the particle production. The availability of future experimental data at other different energies for p+p, p+Pb and Pb+Pb collisions at the LHC would help in a better understanding of this observation.

\item We have found that for p to $\pi$ and $\Lambda$ to $\rm{K}_{s}^0$ ratios, the AMPT-Default version is more closer to the experimental data than that of AMPT-SM specially for $p_{T} >$ 1 GeV/c. This may be due to the coalescence mechanism involved in AMPT-SM, which affects the particle production at intermediate-$p\rm{_T}$.

\item For p to $\phi$ ratio, the AMPT-SM does a better job compared to AMPT-Default version. The AMPT-SM seems to reproduce the experimental data after $p_{T} \sim$ 1 GeV/c, which is expected by the hydrodynamics-inspired models.

\item It is explicitly observed from these extensive studies that the particle ratios are insensitive to nuclear deformation, at least in the case of Xe+Xe collisions. However, it should also be noted here that the particle spectra are sensitive to nuclear deformation.

\item Thermal model analysis of the particle ratios in Xe+Xe collisions at $\sqrt{s_{\rm NN}}$ = 5.44 TeV using AMPT gives the chemical freeze-out temperature for $(0-10\%)$ centrality, $T_{\rm ch}=154 \pm 8$ MeV. Further, we have found $T_{\rm ch}$ to be independent of collision centrality, whereas the $T_{\rm kin}$ is highly centrality dependent. This goes inline with the earlier findings at the RHIC \cite{Abelev:2008ab} and LHC \cite{Sharma:2018jqf}.

\end{enumerate}
We believe the present exhaustive study of the particle spectra, ratios, and freeze-out criteria would be quite helpful in understanding the Xe+Xe collisions at the LHC energies with nuclear deformation, when the corresponding experimental data will be available.
\section*{Acknowledgements}
The authors acknowledge the financial supports  from  ALICE  Project  No. SR/MF/PS-01/2014-IITI(G)  of  
Department  of  Science  \&  Technology,  Government of India. RR and ST acknowledge the financial support by 
DST-INSPIRE program of Government of India. The authors would like to acknowledge the usage of resources  of the LHC grid computing facility at VECC, Kolkata. Dr. Swatantra K. Tiwari is acknowledged for initial discussions and Dr. 
Zi-Wei Lin for the necessary permission for implementing the nuclear deformation in AMPT. The authors are thankful to Arvind Khuntia for his helps in the thermal model analysis.

\end{document}